\documentclass[12pt,letterpaper]{article}
\usepackage[dvips]{graphicx}

\def\title#1{{\bf\Large #1}}
\def\sec#1{\vspace{1.00\baselineskip} \noindent {\bf\large #1}
\vspace{0.25\baselineskip}}

\def\d{\dagger}
\def\<{\langle}
\def\>{\rangle}
\def\tsum{{\textstyle\sum}}

\def\d{\dagger}

\def\ba{\begin{eqnarray}}
\def\ea{\end{eqnarray}}
\def\be{\begin{equation}}
\def\ee{\end{equation}}

\begin{document}


\begin{center}

\vspace*{2.0\baselineskip}

\title{Squeezed states produced by modulation interaction \\
and phase conjugation in fibers} \\

\vspace{1.0\baselineskip}

C. J. McKinstrie \\ {\it\small Bell Laboratories, Alcatel--Lucent, Holmdel, New
Jersey 07733} \\

\vspace{0.50\baselineskip}

Abstract \\

\parbox[t]{5.5in}{\small Number-state expansions are derived for the
squeezed states produced by four-wave mixing (modulation interaction
and phase conjugation) in fibers. These expansions are valid for
arbitrary pump-induced coupling and dispersion-induced mismatch
coefficients. To illustrate their use, formulas are derived for the
associated field-quadrature and photon-number variances and
correlations.}

\end{center}

\newpage

\sec{1. Introduction}

Parametric devices based on four-wave mixing (FWM) in fibers can
generate photon pairs for quantum communication experiments
\cite{sha01,fan07}. Three different types of FWM are illustrated in
Fig. 1. Modulation interaction (MI) is the degenerate process in
which two photons from the same pump are destroyed, and signal and
idler (sideband) photons are created ($2\pi_p \rightarrow \pi_s +
\pi_i$, where $\pi_j$ represents a photon with frequency
$\omega_j$). Inverse MI is the degenerate process in which two
photons from different pumps are destroyed and two signal photons
are created ($\pi_p + \pi_q \rightarrow 2\pi_s$). Phase conjugation
(PC) is the nondegenerate process in which two different pump
photons are destroyed and two different sideband photons are created
($\pi_p + \pi_q \rightarrow \pi_s + \pi_i$). These processes are
reviewed in \cite{mck02,mck04b,mck04c,mck06a}.
\begin{figure}[h!]
\centerline{\includegraphics[height=1.0in]{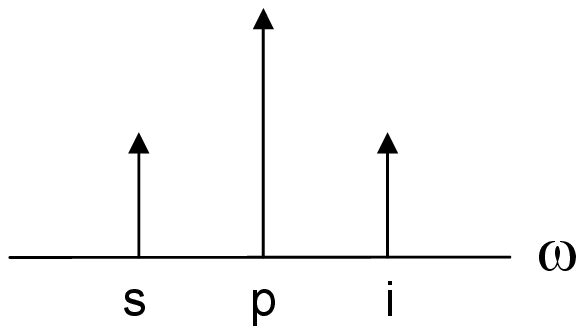}
\hspace{0.2in}
\includegraphics[height=1.0in]{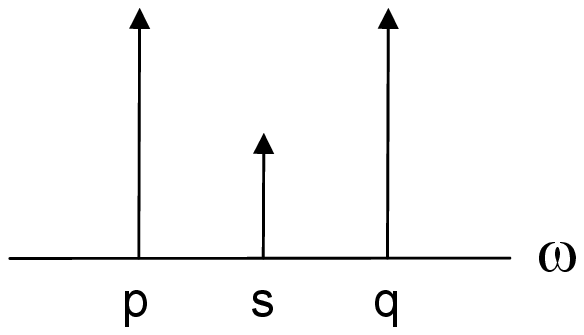}} \vspace{0.14in}
\centerline{\includegraphics[height=1.0in]{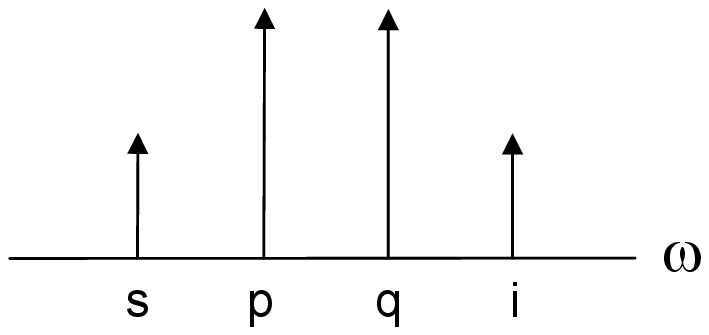}
\hspace{0.2in}
\includegraphics[height=1.0in]{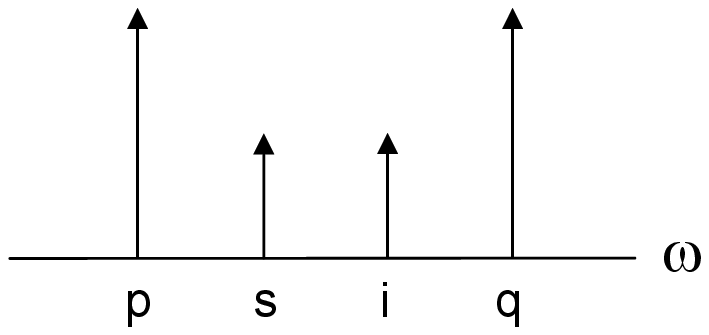}} \caption{Frequency
diagram for ($a$) modulation interaction, ($b$) inverse modulation
interaction, ($c$) outer-band phase conjugation and ($d$) inner-band
phase conjugation in a fiber. Long arrows denote strong pumps ($p$
and $q$), whereas short arrows denote weak sidebands ($s$ and $i$).}
\end{figure}

The evolution of an optical system is governed by the (spatial)
Schr\"odinger equation
\be d_z|\psi\> = iH|\psi\>, \label{1.1} \ee
where $z$ is distance, $d_z = d/dz$, $|\psi\>$ is the state vector,
and the Hamiltonian $H$ depends on the creation and destruction
operators of the interacting modes ($a_j^\d$ and $a_j$,
respectively, where $\d$ denotes a hermitian conjugate). The
solution of Eq. (\ref{1.1}) can be written in the input--output form
\be |\psi(z)\> = U(z)|\psi(0)\>, \label{1.2} \ee
where the evolution operator $U(z) = \exp(iHz)$ is unitary. Notice
that $U(-z) = U^\d(z)$. In the Schr\"odinger picture, the mode
operators are constants.

In the Heisenberg picture, one defines the mode operators $a_j(z) =
U^\d(z)a_j(0)U(z)$, which evolve according to the (spatial)
Heisenberg equations
\be d_za_j = i[a_j,H], \label{1.3} \ee
where $[\ ,\ ]$ denotes a commutator. In the small-signal
(undepleted-pump) regime, the Hamiltonian depends quadratically on
the sideband operators, so Eqs. (\ref{1.3}) are linear in these
operators. Hence, their solutions can be written in the
input--output form
\be a_j(z) = \tsum_k [\mu_{jk}(z)a_k(0) + \nu_{jk}(z)a_k^\d(0)],
\label{1.4} \ee
where $\mu_{jk}(z)$ and $\nu_{jk}(z)$ are transfer functions.

Any measurable quantity associated with mode $j$ can be written as
the expectation value of some function of the mode operator and its
conjugate. Common examples are the field quadrature and photon
number. Let $\<\ \>$ denote an expectation value and $F$ be any
function that has a Taylor expansion. Then
\be \<F(a_j)\> = \<\psi(z)|F[a_j(0)]|\psi(z)\> =
\<\psi(0)|F[a_j(z)]|\psi(0)\>. \label{1.4} \ee
In the Schr\"odinger picture [first part of Eq. (\ref{1.4})], the
state vector evolves and the mode operator is constant, whereas in
the Heisenberg picture [second part of Eq. (\ref{1.4})] the state
vector is constant and the mode operator evolves. Expectation values
of output quantities (which involve one or more operators) can be
calculated using either picture.

To model photon-generation experiments, one needs to determine the
probabilities of measuring different numbers of photons. In this
context, the Schr\"odinger picture is preferable. In this report,
number-state expansions are derived for the squeezed states produced
by (inverse) MI and PC in fibers. These expansions generalize the
standard results \cite{bar97,lou00}, which do not include the
effects of fiber dispersion. As written, they apply to parametric
processes driven by continuous-wave pumps. However, by defining
suitable superposition (Schmidt) modes \cite{law00,mck08b}, one can
also apply them to parametric processes driven by pulsed pumps.


\sec{2. One-mode squeezed state}

The inverse MI in a fiber is governed by the Hamiltonian
\be H = \delta a_s^\d a_s + [\gamma(a_s^\d)^2 + \gamma^*a_s^2]/2,
\label{2.1} \ee
where $a_s$ is the destruction operator of the signal mode, $\delta$
is the mismatch coefficient and $\gamma$ is the coupling
coefficient. Formulas for these coefficients, which involve the
fiber dispersion and nonlinearity coefficients, and the the pump
amplitudes, are stated in \cite{mck04b,mck06a}.

By combining Eqs. (\ref{1.3}) and (\ref{2.1}), one obtains the
evolution equation
\be d_za_s = i\delta a_s + i\gamma a_s^\d . \label{2.2} \ee
The solution of this equation can be written in the input--output
form
\be a_s(z) = \mu(z)a_s(0) + \nu(z)a_s^\d(0), \label{2.3} \ee
where the transfer functions
\ba \mu(z) &= &\cos(kz) + i\delta\sin(kz)/k, \label{2.4} \\
\nu(z) &= &i\gamma\sin(kz)/k \label{2.5} \ea
and the inverse-MI wavenumber $k = (\delta^2 - |\gamma|^2)^{1/2}$.
Equations (\ref{2.4}) and (\ref{2.5}) are based on the assumption
that $k$ is real (and the MI is stable). If $k = i\kappa$ is
imaginary (and the MI is unstable), $\cos(kz)$ is replaced by
$\cosh(\kappa z)$ and $\sin(kz)/k$ is replaced by $\sinh(\kappa
z)/\kappa$. Notice that $\mu(-z) = \mu^*(z)$ and $\nu(-z) =
-\nu(z)$. The transfer functions also satisfy the auxiliary equation
\be |\mu|^2 - |\nu|^2 = 1. \label{2.6} \ee

By definition, the one-mode squeezed state produced by inverse MI is
$U(z)|0\>$, where $U(z) = \exp(iHz)$ and $|0\>$ is the one-mode
vacuum state. One can facilitate the calculation of the output state
by rewriting $U$ in normally-ordered form. To do this, one writes
\be H = \gamma K_+ + 2\delta K_3 + \gamma^*K_- - \delta/2,
\label{2.11} \ee
where the operators
\be K_+ = (a_s^\d)^2/2, \ \ K_- = a_s^2/2, \ \ K_3 = (a_s^\d a_s +
a_sa_s^\d)/4. \label{2.12} \ee
These operators satisfy the angular-momentum-like commutation
relations $[K_+,K_-] = -2K_3$ and $[K_3,K_\pm] = \pm K_\pm$. By
using a standard operator-ordering theorem, which is proved in the
Appendix, one finds that
\be \exp(iHz) = \exp(\gamma_+K_+)\exp(\gamma_3K_3)\exp(\gamma_-K_-),
\label{2.13} \ee
where the auxiliary functions
\ba \gamma_+(z) &= &i\gamma\sin(kz)/[k\cos(kz) - i\delta\sin(kz)], \label{2.14} \\
\gamma_-(z) &= &i\gamma^*\sin(kz)/[k\cos(kz) - i\delta\sin(kz)], \label{2.15} \\
\gamma_3(z) &= &-2\log[\cos(kz) - i\delta\sin(kz)/k] \label{2.16}
\ea
and the (inconsequential) phase factor $\exp(-i\delta z/2)$ was omitted. It
follows from Eq. (\ref{2.13}), and the identities $K_\pm^\d = K_\mp$ and
$K_3^\d = K_3$, that $\gamma_+(-z) = \gamma_-^*(z)$, $\gamma_-(-z) =
\gamma_+^*(z)$ and $\gamma_3(-z) = \gamma_3^*(z)$. The auxiliary functions
satisfy these conditions. By comparing Eqs. (\ref{2.14})--(\ref{2.16}) to Eqs.
(\ref{2.4}) and (\ref{2.5}), one obtains the compact formulas
\be \gamma_+ = \nu/\mu^*, \ \ \gamma_- = -\nu^*/\mu^*, \ \ \gamma_3
= -2\log(\mu^*). \label{2.17} \ee
Hence, if the input is the vacuum state, the output is the squeezed
state
\be |\psi\> = {1 \over (\mu^*)^{1/2}} \sum_{n=0}^\infty \biggl({\nu
\over \mu^*}\biggr)^n {[(2n)!]^{1/2} \over 2^n n!} |2n\>,
\label{2.18} \ee
where the basis vectors $|2n\> = (a_s^\d)^{2n}|0\>/[(2n)!]^{1/2}$.
Notice that each eigenstate contains an even number of photons.

For the special case in which $\delta = 0$ (maximal exponential
growth),
\be \mu = \cosh(|\gamma|z), \ \ \nu =
i\gamma\sinh(|\gamma|z)/|\gamma|, \ \ \nu/\mu^* =
i\gamma\tanh(|\gamma|z)/|\gamma| \label{2.19} \ee
and Eq. (\ref{2.18}) reduces to the standard result
\cite{bar97,lou00}. (To verify this statement, use the substitution
$i\gamma z = -se^{i\theta}$.) For the complementary case in which
$\delta = |\gamma|$ (transitional linear growth),
\be \mu = 1 + i|\gamma|z, \ \ \nu = i\gamma z, \ \ \nu/\mu^* =
i\gamma z/(1 - i|\gamma|z) \label{2.20} \ee
and Eq. (\ref{2.18}) reduces to the result of \cite{mck08a}. (Use
the substitution $|\gamma|z = z$.)

It is customary (and easy) to calculate the moments of $a^\d$ and
$a$ using the Heisenberg picture. However, I will calculate the
lower-order moments using the Schr\"odinger picture, to check Eq.
(\ref{2.18}) and illustrate its use. The zeroth-order moment is
$\<\psi|\psi\>$. Let $c_n$ be the coefficient of $|2n\>$ in Eq.
(\ref{2.18}) and $P_{2n} = |c_n|^2$ be the probability of a
$2n$-photon state. Then $|\mu|P_{2n} = x^{2n} (2n)! /4^n(n!)^2$,
where $x = |\nu/\mu|$. It follows from Eq. (\ref{2.6}) and the
identity
\be S(x) = {1 \over (1 - x^2)^{1/2}} = \sum_{n=0}^\infty
{x^{2n}(2n)! \over 4^n (n!)^2} \label{2.31} \ee
that $\<\psi|\psi\> = 1$: The state vector (\ref{2.18}) is
normalized.

The field quadrature
\be q_s = (a_s^\d e^{i\phi_l} + a_se^{-i\phi_l})/2^{1/2},
\label{2.32} \ee
where $\phi_l$ is the local-oscillator phase, and the quadrature deviation
$\delta q_s = q_s - \<q_s\>$. The first-order moment
\be \< q_s\> = 0, \label{2.33} \ee
because $a^\d |\psi\>$ and $a|\psi\>$ contain only odd-number
states, whereas $\<\psi|$ contains only even-number states. The
expectation values (means) of all the odd moments are zero, for the
same reason. To calculate the quadrature variance $\<\delta q_s^2\>$
(which equals $\<q_s^2\>$), one needs to calculate the inner
products of $\<\psi|$ and
\ba (a_s^\d)^2|\psi\> &= & (\mu^*/\nu)\sum_{n=1}^\infty 2nc_n|2n\>, \label{2.34} \\
a_s^\d a_s|\psi\> &= & \sum_{n=1}^\infty 2nc_n|2n\>, \label{2.35} \\
a_sa_s^\d |\psi\> &= & \sum_{n=0}^\infty (2n + 1)c_n|2n\>, \label{2.36} \\
a_s^2|\psi\> &= & (\nu/\mu^*)\sum_{n=0}^\infty (2n + 1)c_n|2n\>.
\label{2.37} \ea
It follows from Eq. (\ref{2.31}) that
\be |\mu|\sum_{n=0}^\infty 2n P_n = xd_xS(x) = {x^2 \over (1 -
x^2)^{3/2}}. \label{2.38} \ee
By using this result and Eq. (\ref{2.6}) to evaluate the inner
products, one finds that $\<(a_s^\d)^2\> = \mu^*\nu^*$, $\<a_s^\d
a_s\> = |\nu|^2$, $\< a_sa_s^\d\> = |\mu|^2$ and $\< a_s^2\> =
\mu\nu$. Hence, the quadrature variance
\be \<\delta q_s^2\> = (|\mu|^2 + 2|\mu\nu|\cos\theta + |\nu|^2)/2
\label{2.39}, \ee
where the phase difference $\theta = \phi_\mu + \phi_\nu - 2\phi_l$.
The quadrature variance attains its maximum $(|\mu| + |\nu|)^2/2$
when $\theta = 0$ and its minimum $(|\mu| - |\nu|)^2/2$ when $\theta
= \pi$. In the stable regime $|\nu|^2$ is bounded by $\gamma^2/k^2$,
whereas in the unstable regime it is unbounded [Eq. (\ref{2.5})].
The quadrature is squeezed in both regimes. Equation (\ref{2.39}) is
consistent with the results of \cite{mck06a,mck05b}, which were
obtained using the Heisenberg picture. For the special case in which
$\delta = 0$, it reduces to the standard result \cite{bar97,lou00}.

Now define the photon-number operator $n_s = a_s^\d a_s$ and the
number deviation $\delta n_s = n_s - \<n_s\>$. It also follows from
Eq. (\ref{2.31}) that
\be |\mu|\<n_s^m\> = (xd_x)^m S(x). \label{2.40} \ee
By using this result and Eq. (\ref{2.6}), one finds that
\be \<n_s\> = |\nu|^2, \ \ \<n_s^2\> = |\nu|^2(2|\mu|^2 + |\nu|^2),
\ \ \<\delta n_s^2\> = 2|\mu\nu|^2. \label{2.41} \ee
Equations (\ref{2.41}) are consistent with the results of
\cite{mck06a,mck05b}, which were obtained using the Heisenberg
picture. For the special case in which $\delta = 0$, they reduce to
the standard results \cite{bar97,lou00}.


\sec{3. Two-mode squeezed state}

MI and PC in a fiber are governed by the Hamiltonian
\be H = \delta(a_s^\d a_s + a_i^\d a_i) + \gamma a_s^\d a_i^\d +
\gamma^* a_sa_i, \label{3.1} \ee
where $a_j$ is the destruction operator of mode $j$ ($s$ or $i$).
Formulas for the mismatch and coupling coefficients are stated in
\cite{mck02,mck04c}. By combining Eqs. (\ref{1.3}) and (\ref{3.1}),
one obtains the evolution equations
\ba d_za_s &= &i\delta a_s + i\gamma a_i^\d, \label{3.2} \\
d_za_i^\d &= &-i\gamma^*a_s - i\delta a_i^\d. \label{3.3} \ea
The solutions of these equations can be written in the input--output
form
\ba a_s(z) &= &\mu(z)a_s(0) + \nu(z)a_i^\d(0), \label{3.4} \\
a_i^\d(z) &= &\nu^*(z)a_s(0) + \mu^*(z)a_i^\d(0), \label{3.5} \ea
where the transfer functions were defined in Eqs. (\ref{2.4}) and
(\ref{2.5}), and the MI (PC) wavenumber $k = (\delta^2 -
|\gamma|^2)^{1/2}$.

By definition, the two-mode squeezed state produced by MI (PC) is
$U(z)|0,0\>$, where $|0,0\>$ is the two-mode vacuum state. One can
calculate this state by writing
\be H = \gamma K_+ + 2\delta K_3 + \gamma^*K_- - \delta, \label{3.6}
\ee
where the operators
\be K_+ = a_s^\d a_i^\d, \ \ K_- = a_sa_i, \ \ K_3 = (a_s^\d a_i +
a_s a_i^\d)/2. \label{3.7} \ee
These operators also satisfy the commutation relations $[K_+,K_-] = -2K_3$ and
$[K_3,K_\pm] = \pm K_\pm$. By using the aforementioned operator-ordering
theorem, one can rewrite $U$ in the form of Eq. (\ref{2.13}), where the
auxiliary functions were defined in Eqs. (\ref{2.14})--(\ref{2.16}) and the
(inconsequential) phase factor $\exp(-i\delta z)$ was omitted. Hence, if the
input is the vacuum state, the output is the squeezed state
\be |\psi\> = {1 \over \mu^*} \sum_{n=0}^\infty \biggl({\nu \over
\mu^*}\biggr)^n |n,n\>. \label{3.8} \ee
Notice that each eigenstate contains an equal number of signal and
idler photons. For the special cases in which $\delta = 0$ and
$\delta = |\gamma|$, Eq. (\ref{3.8}) reduces to the standard result
\cite{bar97,lou00} and the result of \cite{mck08a}, respectively.

Let $c_n$ be the coefficient of $|n,n\>$ in Eq. (\ref{3.8}) and let
$P_n = |c_n^2|$ be a probability. Then $|\mu|^2P_n = y^n$, where $y
= |\nu/\mu|^2$. It follows from the identity
\be S(y) = {1 \over 1 - y} = \sum_{n=0}^\infty y^n \label{3.11} \ee
that $\<\psi|\psi\> = 1$. By combining Eqs. (\ref{3.8}) and
(\ref{3.11}), one can show that
\be |\mu|^2\<n_j^{m_j}n_k^{m_k}\> = (yd_y)^{m_j+m_k}S(y),
\label{3.12} \ee
where $n_j = a_j^\d a_j$ and $k \neq j$.

The quadrature and photon-number operators of the signal and idler,
and their deviations, are defined in the same way as the signal
operators were defined in Sec. 2. Both quadratures
\be \<q_j\> = 0, \label{3.13} \ee
because $\<\psi|$ contains states with equal numbers of signal and
idler photons, whereas $a_j^\d|\psi\>$ and $a_j|\psi\>$ contain
states with unequal numbers of signal and idler photons: The photon
numbers are unbalanced. Most of the operator moments vanish: Only
powers of $a_j^\d a_j$, $a_ja_j^\d$, $a_j^\d a_k^\d$ and $a_ja_k$
are nonzero. To calculate the quadrature variances $\<\delta
q_j^2\>$ (which equal $\< q_j^2\>$) and correlation $\<\delta
q_j\delta q_k\>$ (which equals $\< q_jq_k\>$), one needs to
calculate the inner products of $\<\psi|$ and
\ba a_j^\d a_k^\d|\psi\> &= &(\mu^*/\nu)\sum_{n=1}^\infty nc_n|n,n\>, \label{3.14} \\
a_j^\d a_j|\psi\> &= &\sum_{n=1}^\infty nc_n|n,n\>, \label{3.15} \\
a_ja_j^\d |\psi\> &= &\sum_{n=0}^\infty (n + 1)c_n|n,n\>, \label{3.16} \\
a_ja_k |\psi\> &= &(\nu/\mu^*)\sum_{n=0}^\infty (n + 1)c_n|n,n\>.
\label{3.17} \ea
It follows from Eq. (\ref{3.12}) that $\<a_j^\d a_j^\d\> =
\mu^*\nu^*$, $\<a_j^\d a_j\> = |\nu|^2$, $\< a_ja_j^\d\> = |\mu|^2$
and $\< a_ja_k\> = \mu\nu$. By combing these results, one finds that
\ba \<\delta q_j^2\> &= &(|\mu|^2 + |\nu|^2)/2, \label{3.18} \\
\<\delta q_s\delta q_i\> &= &|\mu\nu|\cos\theta, \label{3.19} \ea
where the phase difference $\theta = \phi_\mu + \phi_\nu - 2\phi_l$.
Neither of the output modes is squeezed by itself. (The quadrature
variances are phase independent.) Instead, squeezing is manifested
as a quadrature correlation, which strengthens with distance. It
also follows from Eq. (\ref{3.12}) that
\be \<n_j\> = |\nu|^2, \ \ \<n_j^2\> = |\nu|^2(|\mu|^2 + |\nu|^2) =
\<n_jn_k\>. \label{3.20} \ee
In turn, it follows from Eqs. (\ref{3.20}) that
\be \<\delta n_j^2\> = |\mu\nu|^2 = \<\delta n_j\delta n_k\>.
\label{3.21} \ee
Equations (\ref{3.18})--(\ref{3.21}) are consistent with the results
of \cite{mck04c,mck05b}, which were obtained using the Heisenberg
picture. For the special case in which $\delta = 0$, they reduce to
the standard results \cite{bar97,lou00}. Further analysis shows that
$\<(n_j - n_k)^m\> = 0$: The sideband photon-numbers are perfectly
correlated, as implied by Eq. (\ref{3.8}).


\sec{4. Decomposition of a two-mode squeezed state}

It was stated in Sec. 1 that one can relate multiple-mode
transformations to one-mode transformations by defining suitable
superposition modes \cite{law00,mck08b}. This statement applies to
the two-mode transformation discussed in Sec. 3. Define the sum and
difference modes
\be a_\pm = (a_s \pm a_i)/2^{1/2}. \label{4.1} \ee
Then, by making these substitutions in Eq. (\ref{3.1}), one obtains
the alternative Hamiltonian
\ba H &= &\delta a_+^\d a_+ + [\gamma(a_+^\d)^2 + \gamma^*a_+^2]/2
\nonumber \\
& &+\ \delta a_-^\d a_- - [\gamma(a_-^\d)^2 + \gamma^*a_-^2]/2,
\label{4.2} \ea
in which the $+$ and $-$ terms are separate. $H_+$ is identical to
the one-mode Hamiltonian (\ref{2.1}), whereas in $H_-$ the coupling
coefficient $-\gamma$ has the opposite sign. Hence, the two-mode
squeezed state (\ref{3.8}) is the direct product of two one-mode
states of the form (\ref{2.18}), where $\mu_\pm = \mu$ and $\nu_\pm
= \pm\nu$. One can also demonstrate this equivalence directly, by
writing
\ba |\psi\> &= &{1 \over \mu^*} \sum_{n=0}^\infty \biggl({\nu \over
\mu^*}\biggr)^n {(a_s^\d a_i^\d)^n \over n!}|0_s,0_i\> \nonumber \\
&= &{1 \over \mu^*} \sum_{n=0}^\infty \biggl({\nu \over
\mu^*}\biggr)^n {[(a_+^\d)^2 - (a_-^\d)^2]^n \over 2^n n!}|0_+,0_-\> \nonumber \\
&= &{1 \over \mu^*}\sum_{n=0}^\infty \sum_{k=0}^n \biggl({\nu \over
\mu^*}\biggr)^n {(a_+^\d)^{2k}(-1)^{n-k}(a_-^\d)^{2(n-k)} \over 2^n k!(n-k)!} |0_+,0_-\> \nonumber \\
&= &{1 \over \mu^*}\sum_{k=0}^\infty \sum_{l=0}^\infty \biggl({\nu
\over \mu^*}\biggr)^k \biggl({-\nu \over \mu^*}\biggr)^l
{(a_+^\d)^{2k}(a_-^\d)^{2l} \over 2^k k! 2^l l!} |0_+,0_-\>.
\label{4.3} \ea
The last of Eqs. (\ref{4.3}) has the required form.


\sec{5. Summary}

Parametric devices based on modulation interaction (MI) and phase
conjugation (PC) in fibers can generate photon pairs for quantum
communication experiments. In this report, number-state expansions
were derived for the one-mode squeezed state produced by inverse MI
[Eq. (\ref{2.18})], and the two-mode squeezed states produced by MI
and PC [Eq. (\ref{3.8})]. These expansions are valid for arbitrary
pump-induced coupling and dispersion-induced mismatch coefficients.
Hence, they apply to a variety of polarization-dependent parametric
processes driven by continuous-wave pumps in strongly-birefringent,
randomly-birefringent and rapidly-spun fibers. They also apply to
the Schmidt modes that participate in parametric processes driven by
pulsed pumps. To illustrate their use, formulas were derived for the
associated field-quadrature and photon-number variances and
correlations [Eqs. (\ref{2.39}), (\ref{2.41}), (\ref{3.18}),
(\ref{3.19}) and (\ref{3.21})].

\newpage

\sec{Appendix: Operator-ordering theorem}

The main results of this report, Eqs. (\ref{2.18}) and (\ref{3.8}),
were obtained by the use of an operator-ordering theorem (OOT).
Although such theorems are common in the quantum-optics literature
\cite{bar97,lou00}, they are not common in the
optical-communications literature. Consequently, in this appendix
the OOT (\ref{2.13}) will be proved from first principles.

The proof of this OOT relies on the Baker--Campbell--Hausdorff (BCH)
lemma
\begin{equation}
\exp(a)b\exp(-a) = \sum_{n = 0}^\infty [a,b]_n/n!, \label{b1}
\end{equation}
where $a$ and $b$ are operators, and the $n$th-order commutator
$[a,b]_n$ is defined recursively: $[a,b]_0 = b$, $[a,b]_1 = [a,b]$
and $[a,b]_n = [a,[a,b]_{n-1}]$. There are two ways to prove this
lemma. The first (direct) way is to expand both sides of Eq.
(\ref{b1}) in Taylor series, and equate the coefficients of $a^n$
\cite{mck08a}. The second (elegant) way is to define the function
\be F(x) = \exp(xa)b\exp(-xa).  \label{b2} \ee
It follows from Eq. (\ref{b2}) that $F'(x) = aF - Fa = [a,F]$ and $
F''(x) = a[a,F] - [a,F]a = [a,[a,F]]$, where $F' = dF/dx$. By
extending this sequence, and using the fact that $F(0) = b$, one
finds that
\be F(x) = \sum_{n=0}^\infty [a,b]_nx^n/n! \label{b3} \ee
The BCH lemma is Eq. (\ref{b3}), with $x = 1$.

Equation (\ref{2.13}) provides a normally-ordered formula for the
Schr\"odinger evolution-operator $\exp(iHz)$, where $H$ is a
Hamiltonian and $z$ is distance. In this report
\be H = \gamma K_+ + 2\delta K_3 + \gamma^*K_-, \ee
where $\delta$ is real, $K_\pm^\d = K_\mp$ and $K_3^\d = K_3$. The
$K$-operators satisfy the commutation relations $[K_+,K_-] = -2K_3$
and $[K_3,K_\pm] = \pm K_\pm$. (Formulas for these operators were
stated in Secs. 2 and 3.) Because one can multiply $K_+$ and $K_-$
by conjugate phase factors without changing the commutation
relations, one can simplify the derivation of the OOT by assuming
that $\gamma$ is real. Define the function
\begin{equation}
G(z) = \exp[i(\gamma K_+ + 2\delta K_3 + \gamma K_-)z]. \label{b7}
\end{equation}
Because the $K$-operators form a closed set under commutation, one
can rewrite Eq. (\ref{b7}) in the normally-ordered form
\begin{equation}
G(z) = \exp[p(z)K_+]\exp[q(z)K_3]\exp[r(z)K_-], \label{b8}
\end{equation}
where $p$, $q$ and $r$ are functions of $z$ (to be determined). It
follows from Eq. (\ref{b7}) that
\begin{equation}
G' = i(\gamma K_+ + 2\delta K_3 + \gamma K_-)G, \label{b9}
\end{equation}
where $G' = dG/dz$. Likewise, it follows from Eq. (\ref{b8}) that
\begin{equation}
G' = (p'K_+ + q'e^{pK_+}K_3e^{-pK_+} +
r'e^{pK_+}e^{qK_3}K_-e^{-qK_3}e^{-pK_+})G. \label{b10}
\end{equation}
By using lemma (\ref{b1}) and the aforementioned commutation
relations, one finds that
\begin{eqnarray}
e^{pK_+}K_3e^{-pK_+} &= &K_3 - pK_+, \label{b11} \\
e^{qK_3}K_-e^{-qK_3} &= &K_-e^{-q}, \label{b12} \\
e^{pK_+}K_-e^{-pK_+} &= &K_- - 2pK_3 + p^2K_+. \label{b13}
\end{eqnarray}
By using these results to simplify Eq. (\ref{b10}), and equating the
coefficients of $K_+$, $K_3$ and $K_-$ in Eqs. (\ref{b9}) and
(\ref{b10}), one obtains the differential equations
\begin{eqnarray}
p' - pq' + p^2(r'e^{-q}) &= &i\gamma, \label{b14} \\
q' - 2p(r'e^{-q}) &= &2i\delta, \label{b15} \\
r'e^{-q} &= &i\gamma, \label{b16}
\end{eqnarray}
respectively. Equations (\ref{b14})--(\ref{b16}) are to be solved,
subject to the boundary (initial) conditions $p(0) = 0$, $q(0) = 0$
and $r(0) = 0$.

By combining Eqs. (\ref{b14})--(\ref{b16}), one obtains the
individual equation
\be p' = i(\gamma - \delta^2/\gamma) + i\gamma(p + \delta/\gamma)^2.
\label{b17} \ee
This equation has the implicit solution
\be \tan^{-1}[(\gamma p + \delta)/\kappa] - \tan^{-1}[\delta/\kappa]
= i\kappa z, \label{b18} \ee
where the parameter $\kappa = (\gamma^2 - \delta^2)^{1/2}$. By
inverting Eq. (\ref{b18}), one obtains the explicit solution
\be p(z) = i\gamma\sinh(\kappa z)/[\kappa\cosh(\kappa z) -
i\delta\sinh(\kappa z)]. \ee
It is easy to verify that
\ba q(z) &= &-2\log[\cosh(\kappa z) - i\delta\sinh(\kappa z)/\kappa], \label{b19} \\
r(z) &= &i\gamma\sinh(\kappa z)/[\kappa\cosh(\kappa z) -
i\delta\sinh(\kappa z)] \label{b20}
\end{eqnarray}
are the solutions of Eqs. (\ref{b15}) and (\ref{b16}), respectively.
To allow for complex $\gamma$, one replaces $\gamma$ by $\gamma^*$
in Eq. (\ref{b20}) and $\gamma^2$ by $|\gamma|^2$ in the formula for
$\kappa$. These results are consistent with the formulas for
$\gamma_\pm$ and $\gamma_3$ [Eqs. (\ref{2.14})--(\ref{2.16})].

\end{document}